\documentclass[aps,preprint,tightenlines,superscriptaddress,showpacs]{revtex4}
\usepackage{graphicx}
\usepackage{epsfig}
\usepackage{dcolumn}
\usepackage{bm}
\usepackage{color}
\usepackage{overpic}

\newcommand{\xyz}{\rm XYZ}
\newcommand{\x}{X(3872)}
\newcommand{\y}{Y(4260)}

\newcommand{\zc}{Z_c(3900)}
\newcommand{\zcp}{Z_c(4020)}

\newcommand{\BR}{{\cal B}}

\newcommand{\piz}{\pi^0}
\newcommand{\etap}{\eta^{\prime}}
\newcommand{\hc}{h_c}
\newcommand{\pphc}{\pi^+\pi^- h_c}

\newcommand{\psp}{\psi(2S)}
\newcommand{\pspp}{\psi(3770)}
\newcommand{\jpsi}{J/\psi}

\newcommand{\EE}{e^+e^-}
\newcommand{\MM}{\mu^+\mu^-}
\newcommand{\LL}{\ell^+\ell^-}

\newcommand{\pp}{\pi^+\pi^-}

\newcommand{\kk}{K^+K^-}

\newcommand{\ppjpsi}{\pi^+\pi^- J/\psi}

\newcommand{\beq}{\begin{equation}}
\newcommand{\eeq}{\end{equation}}
\newcommand{\bitm}{\begin{itemize}}
\newcommand{\eitm}{\end{itemize}}
\begin{document}

%************************************************************
%%\preprint{} \preprint{\vbox{ \hbox{   }
%%                        \hbox{Version 0}
%%                        }}

\title{%%\quad\\[1.0cm]
\boldmath Study of the $\xyz$ states at the BESIII~\footnote{Talk
at the ``International Workshop on Physics at Future High
Intensity Collider \@ 2--7~GeV in China,'' January 13--16, 2015,
University of Science and Technology of China (USTC), Hefei,
China.}}

\author{Chang-Zheng Yuan}
\email{yuancz@ihep.ac.cn}
 \affiliation{Institute of High Energy Physics, Chinese
Academy of Sciences, Beijing 100049, China}
 \collaboration{for the BESIII Collaboration}

\date{August 7, 2015}

\begin{abstract}

With its unique data samples at energies of 3.8--4.6~GeV, the
BESIII experiment made a significant contribution to the study of
charmonium and charmonium-like states, i.e., the $\xyz$ states. We
review the results for observations of the $\zc$ and $\zcp$
states, the $X(3872)$ in $\EE$ annihilation, and charmonium
$\psi(1\,^3D_2)$ state, as well as measurements of the
cross-sections of $\omega\chi_{cJ}$ and $\eta\jpsi$,  and the
search for $\EE\to \gamma \chi_{cJ}$ and $\gamma Y(4140)$. We also
present  data from BESIII that may further strengthen the study of
the $\xyz$ and conventional charmonium states, and discuss
perspectives on future experiments.

\end{abstract}

\pacs{14.40.Rt, 14.40.Pq, 13.25.Gv, 13.20.Gd, 13.66.Bc}

\maketitle

\section{\boldmath Introduction}

Many charmonium and charmonium-like states were discovered at
$B$-factories in the first decade of the 21st century~\cite{PBFB}.
Whereas some of these are good charmonium candidates, as predicted
in different models, many states have exotic properties, which may
indicate that exotic states, such as multi-quark, molecule,
hybrid, or hadron-quarkonium, have been observed~\cite{review}.

The BESIII experiment~\cite{bes3} at the BEPCII storage ring
started its first collisions in the tau-charm energy region in
2008. After a few years running at suitable energies for its
well-defined physics programs~\cite{BESIII_YB}, i.e., at $\jpsi$
and $\psp$ peaks in 2009 and the $\pspp$ peak in 2010 and 2011,
the BESIII experiment started to collect data for the study of the
$\xyz$ particles, which were not described in the Yellow
Book~\cite{BESIII_YB}.

As the design center-of-mass (c.m.) energy of the BEPCII was
2.0--4.2~GeV, there were not many options for data samples relevant to
the $\xyz$-related physics. BESIII took its first data sample at
the peak of $\psi(4040)$ in May 2011, with the aim of searching for the well-known
$X(3872)$ in the $\psi(4040)$ radiative transition and possibly the
excited $P$-wave charmonium spin-triplet states in similar
transitions. This sample is about 0.5~fb$^{-1}$, which is limited
by the one-month running time left after the $\pspp$ data taken
in the 2010--2011 run. Data were not collected at the
 $\psi(4170)$ peak because the CLEO-c experiment had already
collected a sample of about 0.6~fb$^{-1}$ for the study of $D_s$
decays, which could be used for similar studies.

The upgrade of BEPCII's LINAC in summer 2012 increased the highest
beam energy from 2.1 to 2.3~GeV, making it possible to collect
data at higher c.m. energies (up to 4.6~GeV). This made data
collection possible at almost all known vector states, including
$Y(4260)$, $Y(4360)$, $\psi(4415)$, and (marginally) $Y(4660)$.

The data collected at a c.m. energy of $\sqrt{s}=4.26$~GeV turned
out to be very fruitful. One month's data of 525~pb$^{-1}$ (from
December 14, 2012 to January 14, 2013) produced observations of
the charged charmonium-like state $\zc$~\cite{zc3900}, resulting
in changes to the data collection plan for the 2012--2013 run.
More data were accumulated at c.m. energies of 4.26~GeV and then
4.23~GeV. Data from the $Y(4360)$ peak were also obtained in
spring 2013, and data from even higher energies (4.42 and 4.6~GeV)
were recorded in 2014 after a fine scan of the total hadronic
cross-sections between 3.8 and 4.6~GeV at more than 100 energy
points, with a total integrated luminosity of about 800~pb$^{-1}$.

The dedicated data samples for the $\xyz$ study are presented in
Table~\ref{ecm_lum_xyz}, which lists the nominal c.m. energy,
measured c.m. energy, and  integrated luminosity at each energy
point. These data were used for all the analyses presented in this
article.

\begin{table*}[htbp]
 \centering
\caption{\label{ecm_lum_xyz} The measured c.m.
energy~\cite{ecm_gaoq}, integrated luminosity~\cite{lum_songwm} of
each data sample collected for the study of the $\xyz$ states. The
uncertainties on the integrated luminosities are statistical only;
a 1\% systematic uncertainty common to all the data points is not
listed. }
\begin{tabular}{ccr}
\hline
Data sample & c.m. energy~(MeV)  & ${\cal L}$ ($\rm pb^{-1}$) \\
\hline
3810       &   3807.65$\pm$0.10$\pm$0.58            &50.54$\pm$0.03      \\
3900       &   3896.24$\pm$0.11$\pm$0.72            &52.61$\pm$0.03      \\
4009       &   4007.62$\pm$0.05$\pm$0.66            &481.96$\pm$0.01     \\
4090       &   4085.45$\pm$0.14$\pm$0.66            &52.63$\pm$0.03     \\
4190       &   4188.59$\pm$0.15$\pm$0.68            &43.09$\pm$0.03      \\
4210       &   4207.73$\pm$0.14$\pm$0.61            &54.55$\pm$0.03     \\
4220       &   4217.13$\pm$0.14$\pm$0.67            &54.13$\pm$0.03     \\
4230       &   4226.26$\pm$0.04$\pm$0.65            &1091.74$\pm$0.15  \\
4245       &   4241.66$\pm$0.12$\pm$0.73            &55.59$\pm$0.04     \\
4260       &   4257.97$\pm$0.04$\pm$0.66            &825.67$\pm$0.13   \\
4310       &   4307.89$\pm$0.17$\pm$0.63            &44.90$\pm$0.03   \\
4360       &   4358.26$\pm$0.05$\pm$0.62            &539.84$\pm$0.10  \\
4390       &   4387.40$\pm$0.17$\pm$0.65            &55.18$\pm$0.04   \\
4420       &   4415.58$\pm$0.04$\pm$0.72            &1073.56$\pm$0.14  \\
4470       &   4467.06$\pm$0.11$\pm$0.73            &109.94$\pm$0.04   \\
4530       &   4527.14$\pm$0.11$\pm$0.72            &109.98$\pm$0.04   \\
4575       &   4574.50$\pm$0.18$\pm$0.70            &47.67$\pm$0.03    \\
4600       &   4599.53$\pm$0.07$\pm$0.74            &566.93$\pm$0.11   \\
\hline
\end{tabular}
\end{table*}

\section{\boldmath Charged charmonium-like states: $Z_c$s}

The BESIII experiment observed, for the first time, a charged
charmonium-like state close to the $D\bar{D}^*$ threshold
$Z_c(3900)$/$Z_c(3885)$, and a charged charmonium-like state close
to the $D^*\bar{D}^*$ threshold $Z_c(4020)$/$Z_c(4025)$. Their
neutral partners were also observed, confirming their isospins to
be one.

\subsection{\boldmath Observation of  $Z_c(3900)$ and $Z_c(3885)$}

\subsubsection{Observation of  $Z_c(3900)$}

\begin{figure}[htbp]
 \includegraphics[height=5.3cm]{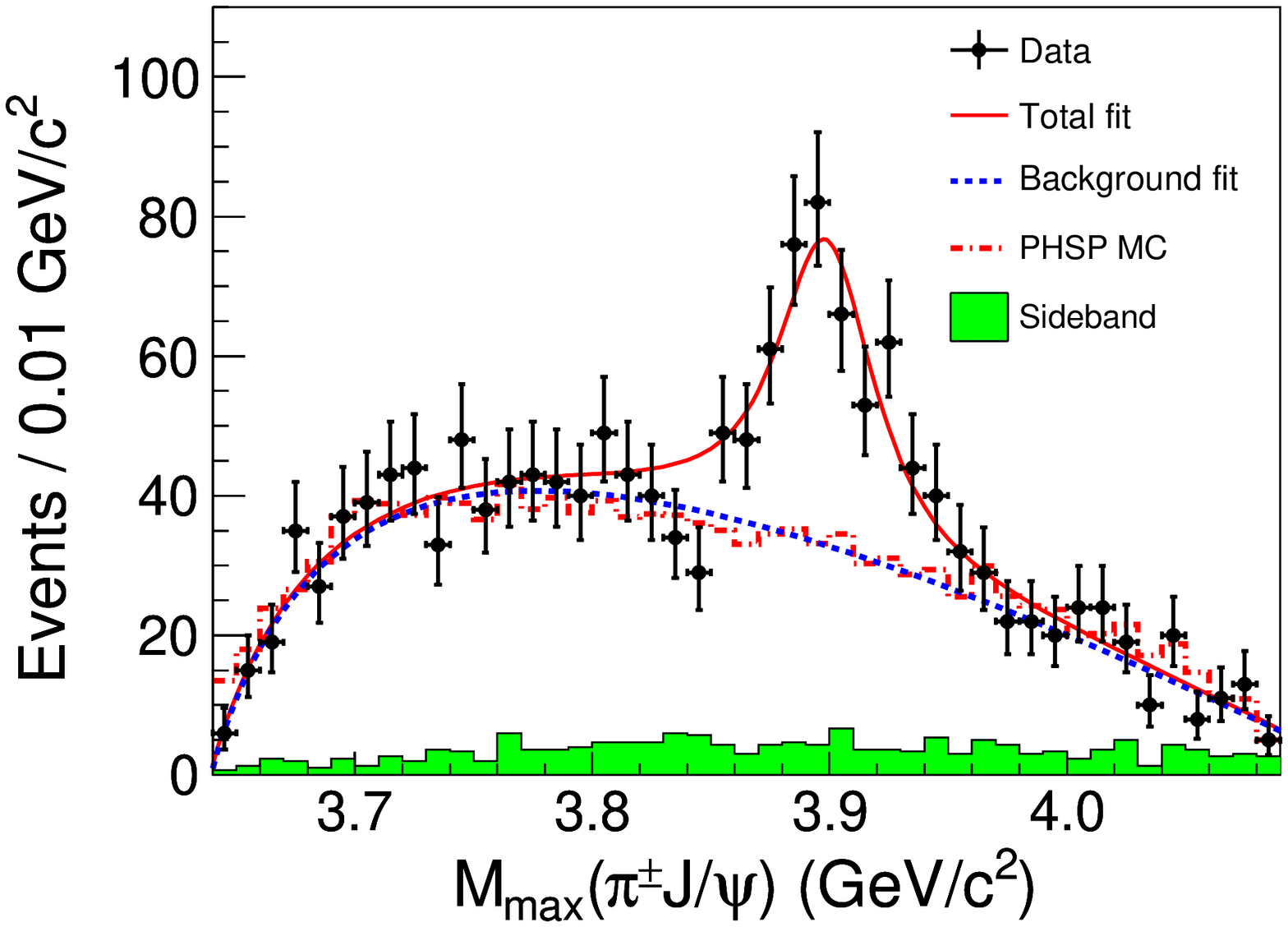}
 \includegraphics[height=5.3cm]{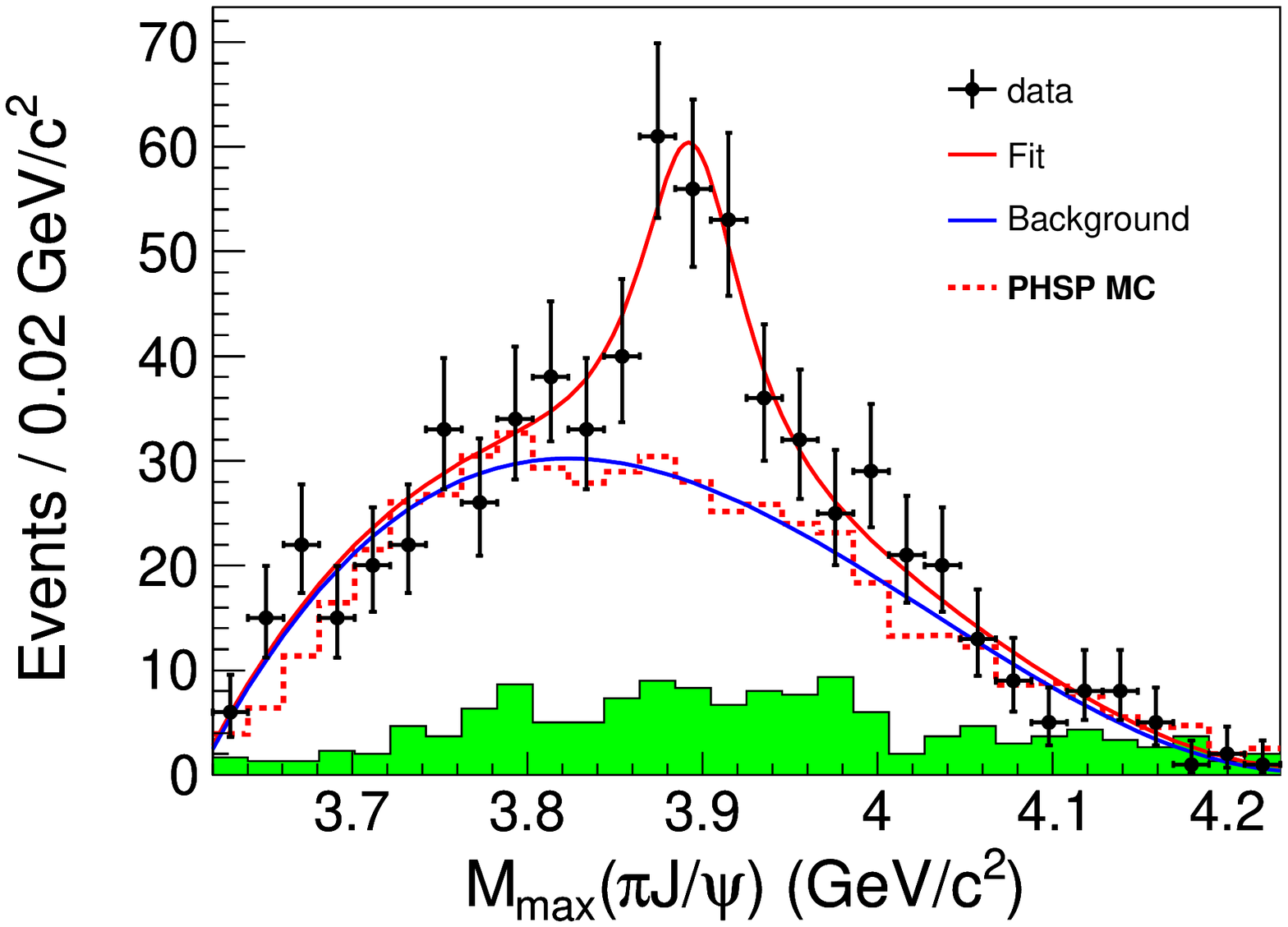}
 \caption{Unbinned maximum likelihood fit to the distribution of
the $M_{\mathrm{max}}(\pi J/\psi)$ (left panel from BESIII and
right panel from Belle). Points with error bars are data, the
curves are the best fit, the dashed histograms are the phase space
distributions and the shaded histograms are the non-$\ppjpsi$
background estimated from the normalized $\jpsi$ sidebands.}
\label{projfit}
\end{figure}

The BESIII experiment studied the $\EE\to \ppjpsi$ process at a
c.m. energy of 4.26~GeV using a 525~pb$^{-1}$ data
sample~\cite{zc3900}. A structure at around 3.9~GeV/$c^2$ was
observed in the $\pi^\pm \jpsi$ mass spectrum with a statistical
significance larger than $8\sigma$, which is referred to as the
$\zc$. A fit to the $\pi^\pm\jpsi$ invariant mass spectrum (see
Fig.~\ref{projfit}), neglecting interference, results in a mass of
$(3899.0\pm 3.6\pm 4.9)~{\rm MeV}/c^2$ and a width of $(46\pm
10\pm 20)$~MeV. The associated production ratio was measured to be
$R=\frac{\sigma(\EE\to \pi^\pm \zc^\mp\to \ppjpsi))}
{\sigma(\EE\to \ppjpsi)}=(21.5\pm 3.3\pm 7.5)\%$.

In the Belle experiments, the cross-section of $\EE\to \ppjpsi$
was measured from 3.8--5.5~GeV using the initial state radiation
(ISR) method. The intermediate states in $\y\to \ppjpsi$ decays
were also investigated~\cite{belley_new}. The $\zc$ state
(referred to as $Z(3900)^+$ in the Belle paper) with a mass of
$(3894.5\pm 6.6\pm 4.5)~{\rm MeV}/c^2$ and a width of $(63\pm
24\pm 26)$~MeV was observed in the $\pi^\pm\jpsi$ mass spectrum
(see Fig.~\ref{projfit}) with a statistical significance larger
than $5.2\sigma$.

%%%% M(max) fit %%%%

The $\zc$ state was confirmed shortly after with CLEO-c data at a
c.m. energy of 4.17~GeV~\cite{seth_zc}, and the mass and width
agreed very well with the BESIII and Belle measurements.

A neutral state $Z_c(3900)^{0}\to \piz\jpsi$ with a significance
of $10.4\sigma$ was observed at BESIII in $e^+e^-\to \pi^0\pi^0
J/\psi$ with c.m. energy ranges from
4.19--4.42~GeV~\cite{zc03900}. The mass and width were measured to
be $(3894.8\pm 2.3\pm 3.2)$~MeV/$c^2$ and $(29.6\pm 8.2\pm
8.2)$~MeV, respectively. This state is interpreted as the neutral
partner of the $Z_c(3900)^\pm$, as it decays to $\pi^0 J/\psi$ and
its mass is close to that of $Z_c(3900)^\pm$. This is in agreement
with the previously reported $3.5\sigma$ evidence for
$Z_c(3900)^{0}$ in the CLEO-c data~\cite{seth_zc}. The measured
Born cross-sections of $e^+e^-\to \pi^0\pi^0 J/\psi$ were about
half of those for $e^+e^-\to \pi^+\pi^- J/\psi$ measured in the
Belle experiment~\cite{belley_new}, which is consistent with the
isospin symmetry expectation.

\subsubsection{Observation of  $Z_c(3885)$}

\begin{figure}[htbp]
\centering
  \includegraphics[width=7.0cm]{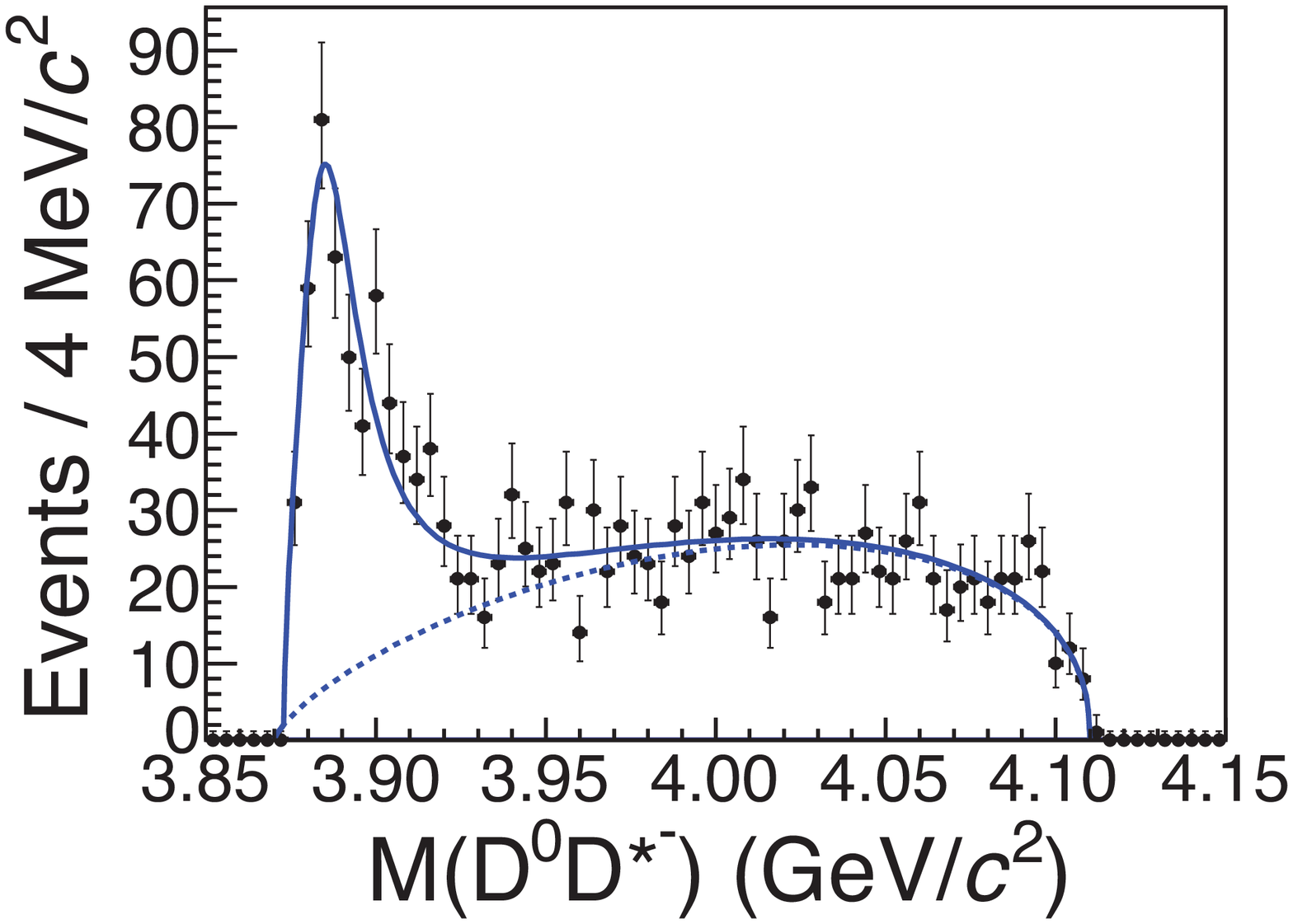}
  \includegraphics[width=7.0cm]{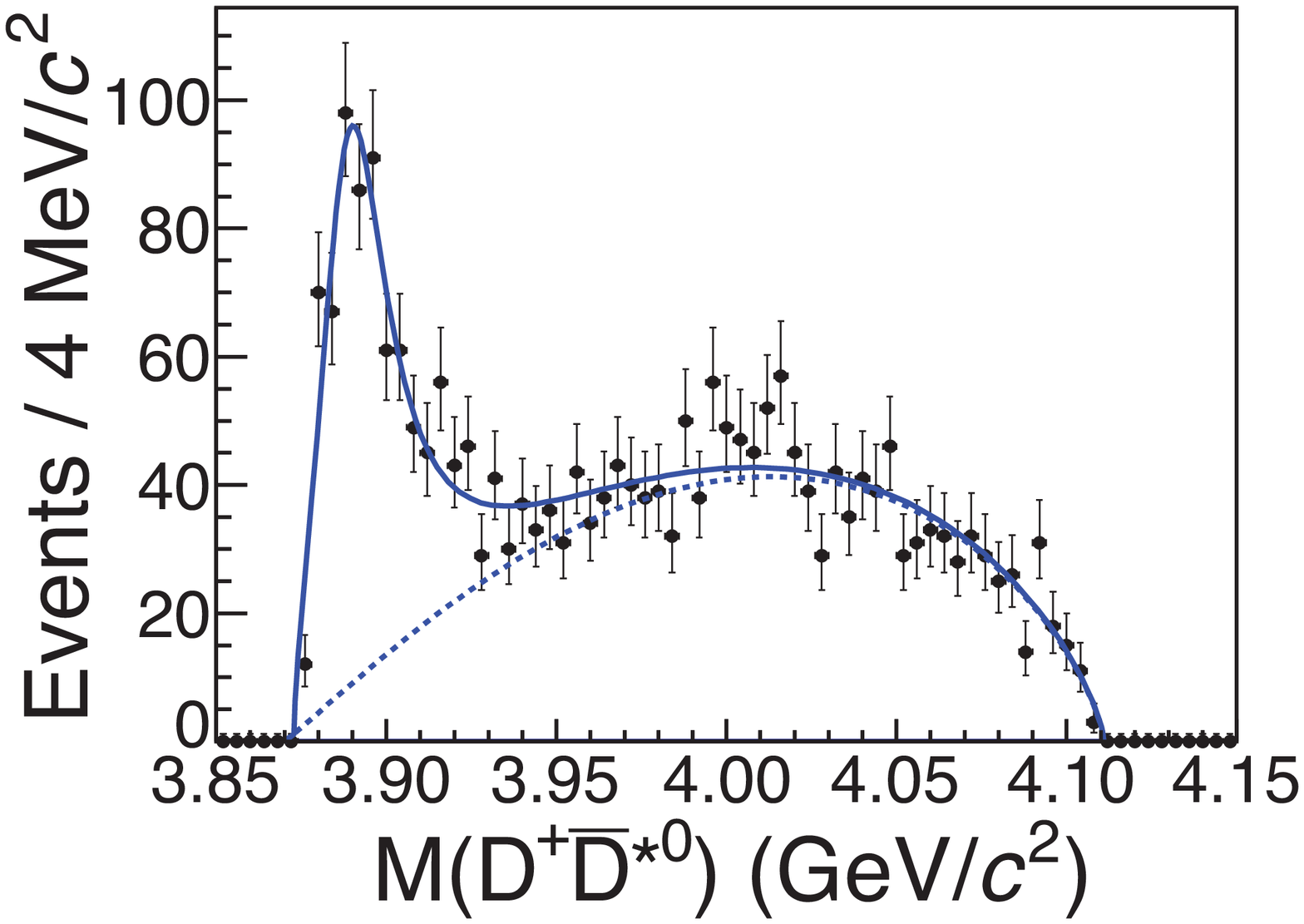}
\caption{The $M(D^0 D^{*-})$ (left) and $M(D^+\bar{D}^{*0})$
(right) distributions for selected events at $\sqrt{s}=4.26$~GeV.
The curves show the best fits. } \label{xuxp}
\end{figure}

The $\zc$ state observed in the $\pi\jpsi$ final state is close to
and above the $D\bar{D}^*$ mass threshold. With the same data
sample at $\sqrt{s}=4.26$~GeV, the BESIII experiment studied
$\EE\to \pi^\pm (D\bar{D}^*)^{\mp}$. A structure (referred to as
$Z_c(3885)$) was observed in the $(D\bar{D}^*)^{\pm}$ invariant
mass distribution~\cite{zc3885}. When fitted to a
mass-dependent-width Breit--Wigner (BW) function, the pole mass
and width were determined to be $(3883.9 \pm 1.5 \pm
4.2)$~MeV/$c^2$ and $(24.8\pm 3.3 \pm 11.0)$~MeV, respectively
(see Fig.~\ref{xuxp}). The angular distribution of the $Z_c(3885)$
system favors a $J^{P}=1^{+}$ assignment for the structure and
disfavors $1^-$ or $0^-$. The production rate was measured to be
$\sigma(\EE \to \pi^{\mp} Z_c(3885)^{\pm})\times
\BR(Z_c(3885)^{\pm}\to (D\bar{D}^*)^{\pm}) =(83.5\pm 6.6 \pm
22.0)$~pb.

An important question is whether $Z_c(3885)$ is the same as
$\zc$~\cite{zc3900,belley_new}. The mass and width of $Z_c(3885)$
are $2\sigma$ and $1\sigma$, respectively, below those of  $\zc$,
as observed by the BESIII and Belle experiments. However, neither
fit considers the possibility of interference with a coherent
non-resonant background, which could shift the results. A
spin-parity quantum number determination for $Z_c(3900)$ would
provide an additional test of this possibility.

Assuming the $Z_c(3885)$ structure is caused by $Z_c(3900)$, we
obtain $\frac{\Gamma(Z_c(3885)\to D\bar{D}^*)}
{\Gamma(Z_c(3900)\to \pi\jpsi)} = 6.2\pm 1.1\pm 2.7$. This ratio
is much smaller than typical values for decays of conventional
charmonium states above the open charm threshold, e.g.,
$\Gamma(\psi(3770)\to D\bar{D})/\Gamma(\psi(3770)\to
\pp\jpsi)=482\pm 84$~\cite{pdg} and $\Gamma(\psi(4040)\to
D^{(*)}\bar{D}^{(*)}) / \Gamma(\psi(4040)\to \eta\jpsi)=192\pm
27$~\cite{4040}. This suggests very different dynamics in the
$\y$-$\zc$ system.

\subsection{\boldmath Observation of  $Z_c(4020)$ and $Z_c(4025)$}

\subsubsection{Observation of $Z_c(4020)$}

\begin{figure}[htbp]
\begin{center}
\includegraphics[width=0.45\textwidth]{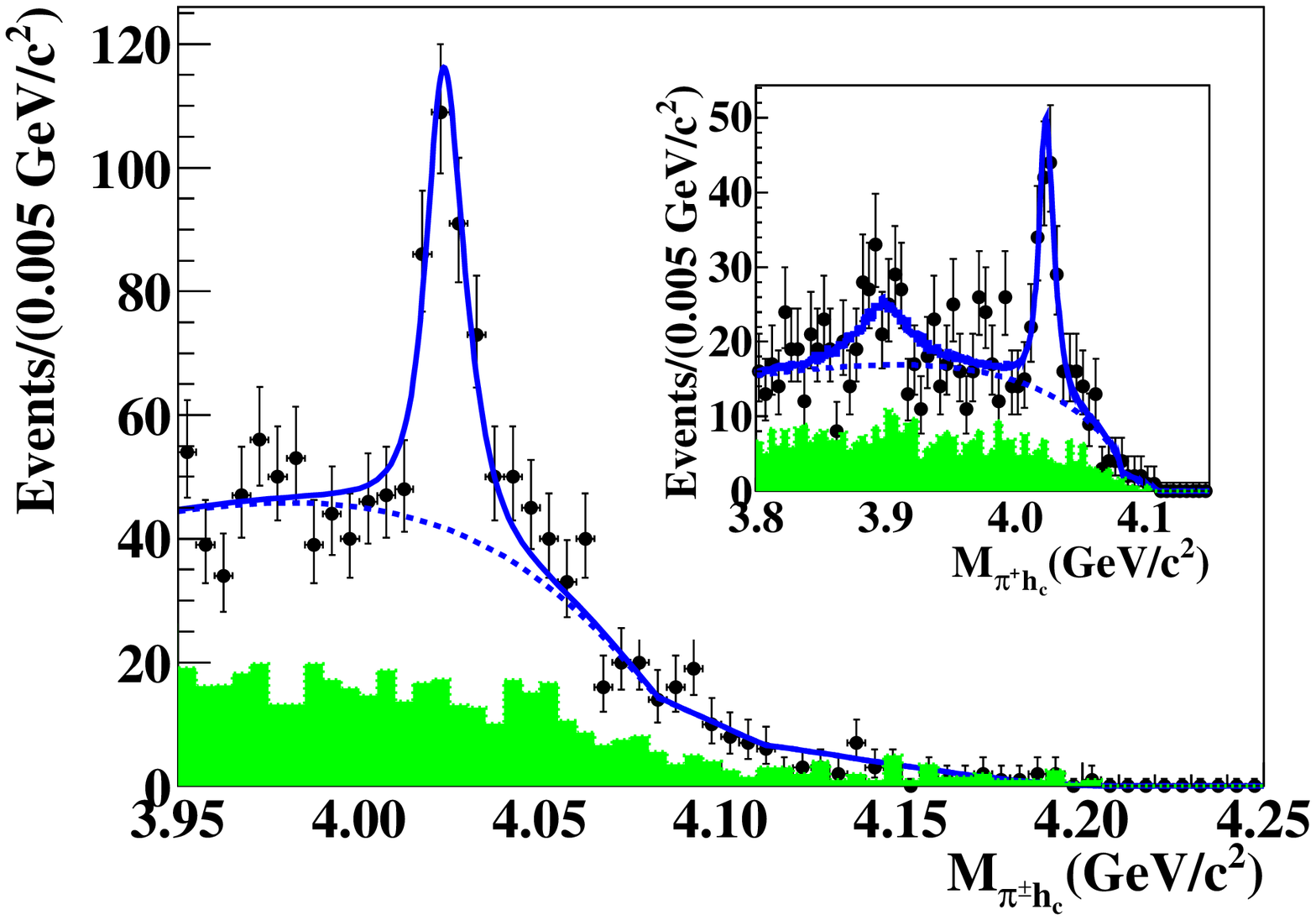}
\includegraphics[width=0.45\textwidth]{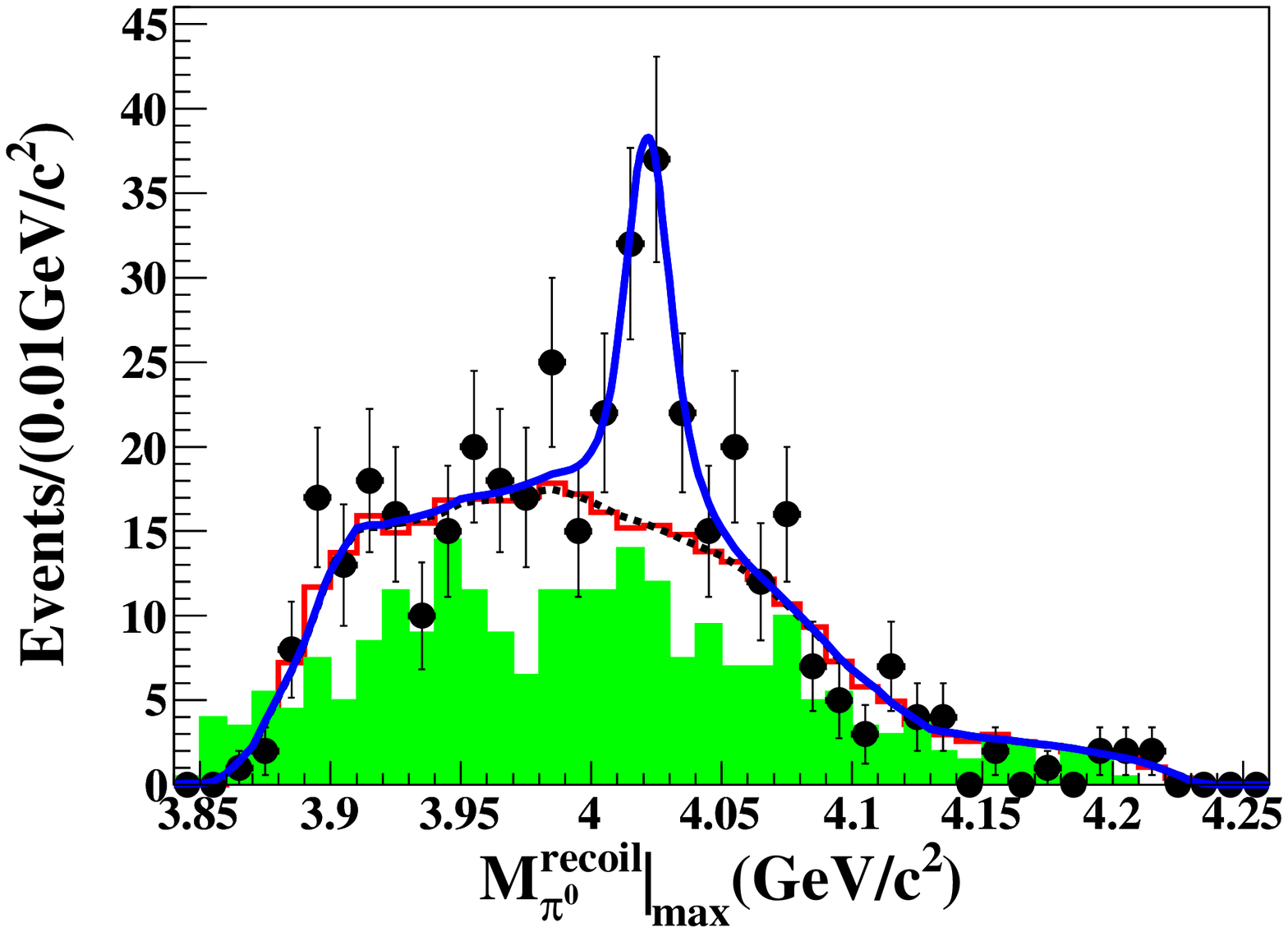}
\caption{Sum of the simultaneous fits to the $M(\pi^\pm h_c)$
(left panel) and $M(\piz h_c)$ (right panel) distributions at
4.23, 4.26, and 4.36~GeV in the BESIII data; the inset in the left
panel shows the sum of the simultaneous fit to the $M_{\pi^+ h_c}$
distributions at 4.23 and 4.26~GeV with $\zc$ and $\zcp$. Dots
with error bars are data; shaded histograms are normalized
sideband background; the solid curves show the total fit, and the
dotted curves the backgrounds from the fit.} \label{1Dfit}
\end{center}
\end{figure}

BESIII measured $\EE\to \pphc$ cross-sections~\cite{zc4020} at
c.m. energies of 3.90--4.42~GeV. Intermediate states were studied
by examining the Dalitz plot of the selected $\pphc$ candidate
events. The $\hc$ signal was selected using $3.518 < M_{\gamma
\eta_c} < 3.538$~GeV/$c^2$, and $\pphc$ samples of 859 events at
4.23~GeV, 586 events at 4.26~GeV, and 469 events at 4.36~GeV were
obtained with purities of~65\%. Although there are no clear
structures in the $\pp$ system, there is clear evidence for an
exotic charmonium-like structure in the $\pi^\pm\hc$ system, as
clearly shown in the Dalitz plot. Figure~\ref{1Dfit}~(left) shows
the projection of the $M(\pi^\pm\hc)$ (two entries per event)
distribution for the signal events, as well as the background
events estimated from normalized $\hc$ mass sidebands. There is a
significant peak at around 4.02~GeV/$c^2$ ($\zcp$), and there are
also some events at around 3.9~GeV/$c^2$ (inset of
Fig.~\ref{1Dfit}~(left)), which could be $\zc$. The individual
datasets at $\sqrt{s}=4.23$, 4.26, and 4.36~GeV show similar
structures.

An unbinned maximum likelihood fit was applied to the
$M(\pi^\pm\hc)$ distribution summed over the 16 $\eta_c$ decay
modes. The data at 4.23, 4.26, and 4.36~GeV were fitted
simultaneously to the same signal function with common mass and
width. Figure~\ref{1Dfit}~(left) shows the fitted results. The
mass and width of  $\zcp$ were measured to be $(4022.9\pm 0.8\pm
2.7)~{\rm MeV}/c^2$ and  $(7.9\pm 2.7\pm 2.6)$~MeV, respectively.
The statistical significance of the $\zcp$ signal was found to be
greater than $8.9\sigma$.

Adding $\zc$ with mass and width fixed to the BESIII
measurements~\cite{zc3900} to the fit results in a statistical
significance of 2.1$\sigma$ (see the inset of
Fig.~\ref{1Dfit}~(left)). At the 90\% confidence level (C.L.), the
upper limits on the production cross-sections are set to
$\sigma(\EE\to \pi^\pm \zc^\mp\to \pphc) <13$~pb at 4.23~GeV and
$<11$~pb at 4.26~GeV. These are lower than those of $\zc\to
\pi^\pm\jpsi$~\cite{zc3900}.

BESIII also observed $\EE\to \piz\piz\hc$ at $\sqrt{s}=4.23$,
4.26, and 4.36~GeV for the first time~\cite{zc0_4020}. The
measured Born cross-sections were about half of those for $\EE\to
\pi^+\pi^-h_c$, which agree with expectations based on isospin
symmetry within systematic uncertainties. A narrow structure with
a mass of $(4023.9\pm 2.2\pm 3.8)$~MeV/$c^2$ (for fitting, the
width was fixed to that measured in the $\EE\to \pi^{+}\pi^{-}h_c$
process~\cite{zc4020} because of low statistics) was observed in
the $\piz\hc$ mass spectrum (Fig.~\ref{1Dfit}~(right)). This
structure is most likely the neutral isospin partner of the
charged $Z_c(4020)$ observed in the $\EE\to \pi^{+}\pi^{-}h_c$
process~\cite{zc4020}. This observation indicates that there are
no anomalously large isospin violations in $\pi\pi h_c$ and $\pi
Z_c(4020)$ systems.

\subsubsection{Observation of  $Z_c(4025)$}

\begin{figure}[htbp]
\centering
\includegraphics[height=6cm]{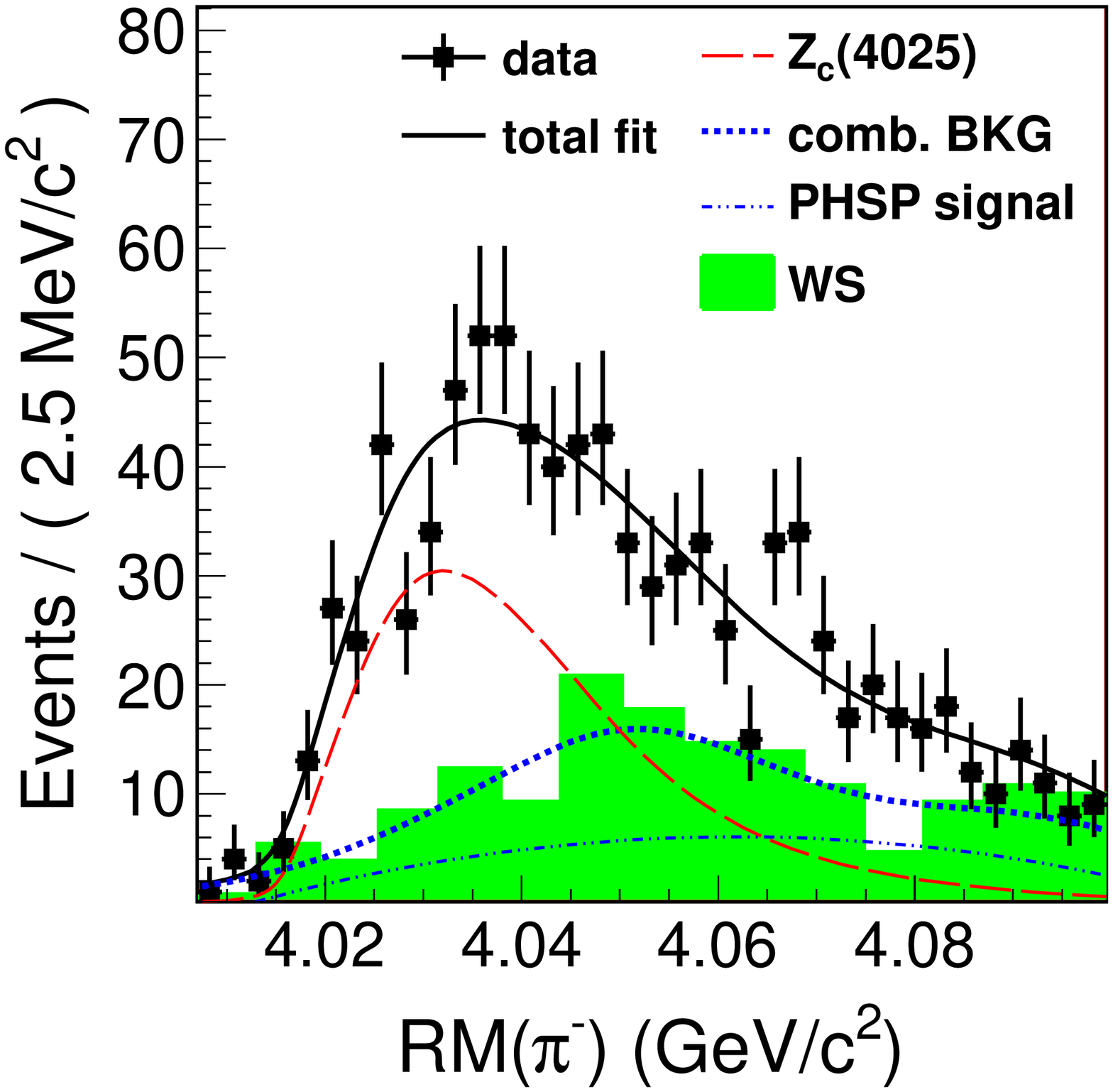}
\includegraphics[height=6cm]{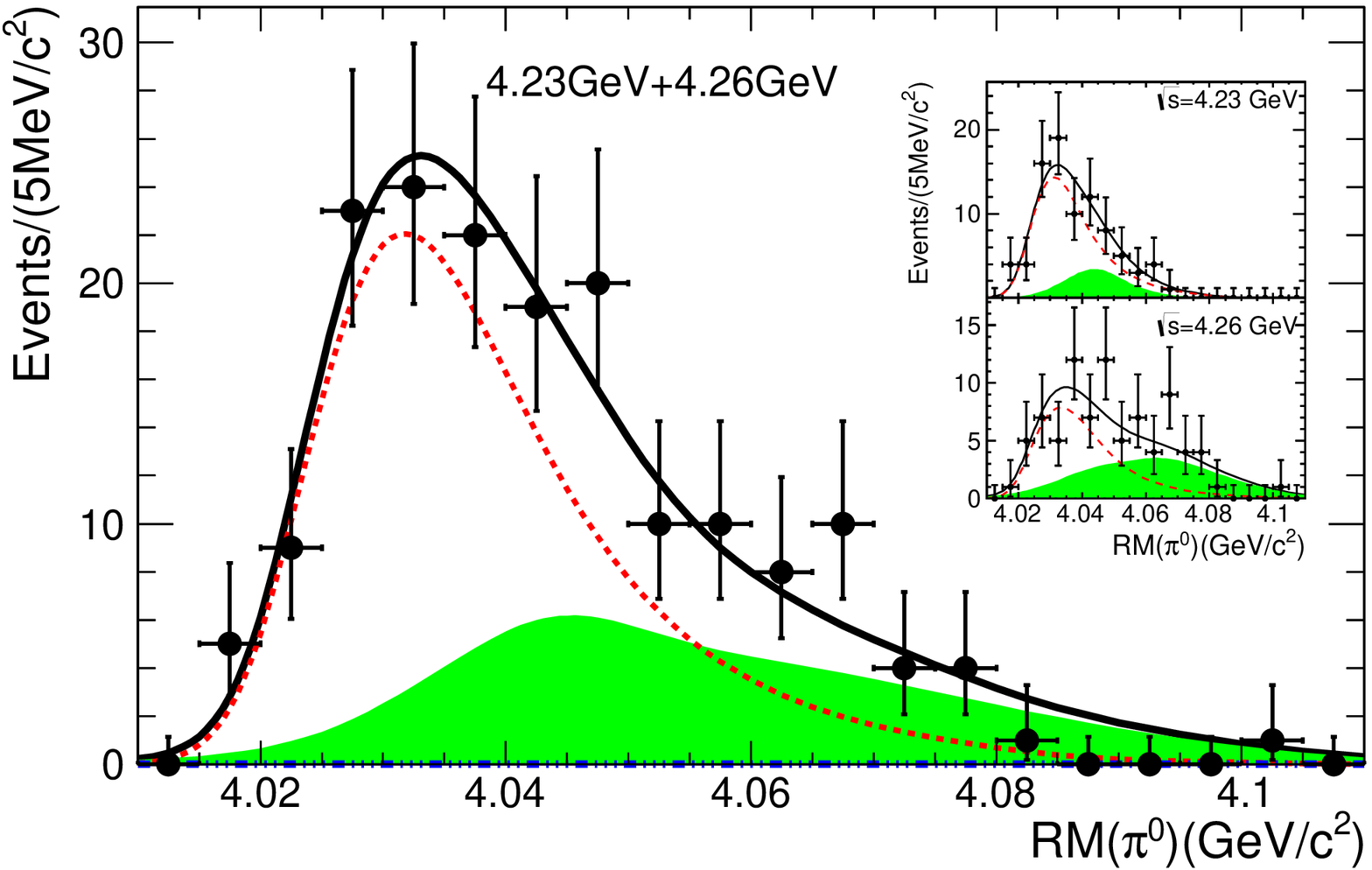}
\caption{Unbinned maximum likelihood fit to the $\pi^\mp$ recoil
mass spectrum (left) in $e^+e^- \to (D^{*} \bar{D}^{*})^{\pm}
\pi^\mp$ at $\sqrt{s}=4.26$~GeV, and to the $\piz$ recoil mass
spectrum (right) in $e^+e^- \to (D^{*} \bar{D}^{*})^{0} \piz$ at
$\sqrt{s}=4.23$ and 4.26~GeV at BESIII. } \label{fig:fit}
\end{figure}

The BESIII experiment also studied the $e^+e^- \to (D^{*}
\bar{D}^{*})^{\pm} \pi^\mp$ process at 4.26~GeV using an
827~pb$^{-1}$ data sample~\cite{zc4025}. Based on a partial
reconstruction technique, the Born cross-section was measured to
be $(137\pm 9\pm 15)$~pb. A structure near the $(D^{*}
\bar{D}^{*})^{\pm}$ threshold in the $\pi^\mp$ recoil mass
spectrum was observed, and this is denoted as $Z_c(4025)$ (see
Fig.~\ref{fig:fit}~(left)). The measured mass and width of the
structure were $(4026.3\pm 2.6\pm 3.7)$~MeV/$c^2$ and $(24.8\pm
5.6\pm 7.7)$~MeV, respectively, from the fit with a constant-width
BW function for the signal. The associated production ratio
$\frac{\sigma(e^+e^-\to Z^{\pm}_c(4025)\pi^\mp \to (D^{*}
\bar{D}^{*})^{\pm} \pi^\mp)}{\sigma(e^+e^-\to (D^{*}
\bar{D}^{*})^{\pm} \pi^\mp)}$ was determined to be $0.65\pm
0.09\pm 0.06$.

Using data at $\sqrt{s}=4.23$ and 4.26~GeV, a structure was
observed in the $\piz$ recoil mass spectrum in the $e^{+}e^{-} \to
D^{*0} \bar{D}^{*0} (D^{*+}{D}^{*-}) \piz$
process~\cite{zc0_4025}. Assuming that the enhancement is due to a
neutral state decaying to $D^{*}\bar{D}^{*}$, the mass and width
of its pole position were determined to be
$(4025.5^{+2.0}_{-4.7}\pm 3.1)$~MeV/$c^2$ and $\Gamma= (23.0\pm6.0
\pm 1.0)$~MeV, respectively (see Fig.~\ref{fig:fit}~(right)). The
Born cross-section $\sigma(e^{+}e^{-} \to Z_c(4025)^0 \piz
\to(D^{*0} \bar{D}^{*0} + D^{*+}{D}^{*-})\piz)$ was measured to be
$(61.6\pm 8.2\pm 9.0)~\rm{pb}$ at 4.23~GeV and $(43.4\pm8.0 \pm
5.4)~\rm{pb}$ at 4.26~GeV. The ratio $\frac{\sigma(\EE \to
Z_c(4025)^0 \piz \to(D^{*} \bar{D}^{*})^0\piz)}{\sigma(\EE\to
Z_c(4025)^+ \pi^-\to(D^{*} \bar{D}^{*})^+\pi^-)}$ is compatible
with unity at $\sqrt{s}=4.26$~GeV, which is expected from isospin
symmetry. In addition, $Z_c(4025)^0$ has a mass and width that are
very close to those of $Z_c(4025)^{\pm}$, which couples to $(D^{*}
\bar{D}^{*})^{\pm}$. Therefore, the observed $Z_c(4025)^0$ state
is a good candidate for the isospin partner of $Z_c(4025)^{\pm}$.

As the $Z_c(4025)$ parameters agree to within 1.5$\sigma$ with
those of $\zcp$, it is very probable that they are the same state.
As the results for $Z_c(4025)^{\pm}$ are only from data at
4.26~GeV, extending the analysis to 4.23~GeV and 4.36~GeV will
probably provide a definite answer.

\section{\boldmath Observation of $\y\to\gamma X(3872)$}
\label{sec:x3872}

BESIII observed $\EE\to \gamma\x\to \gamma \ppjpsi$, with $\jpsi$
reconstructed through its decays into lepton pairs ($\LL=\EE$ or
$\MM$)~\cite{BES3x}.

\begin{figure}
\begin{center}
\includegraphics[height=5cm]{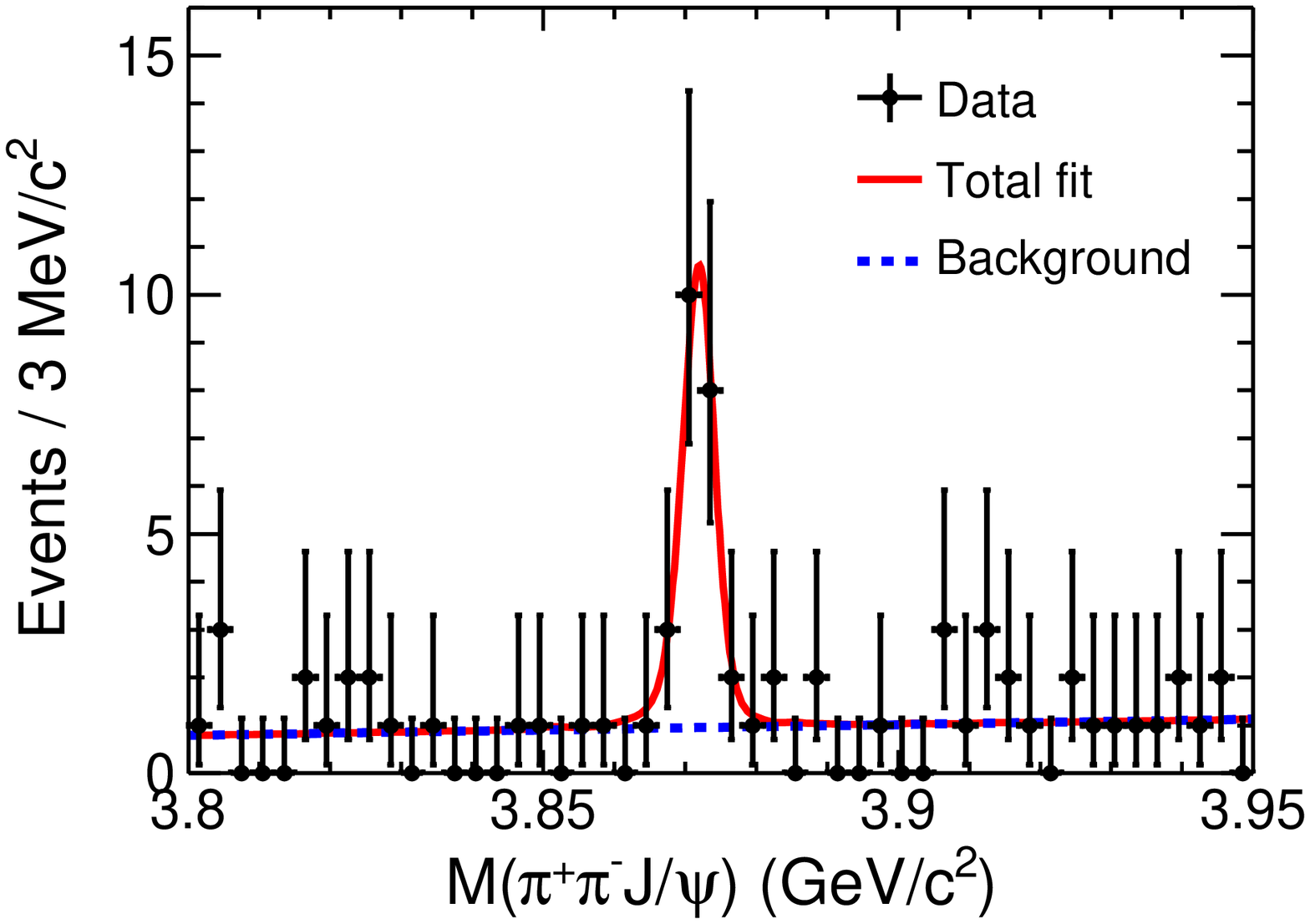}
\includegraphics[height=5cm]{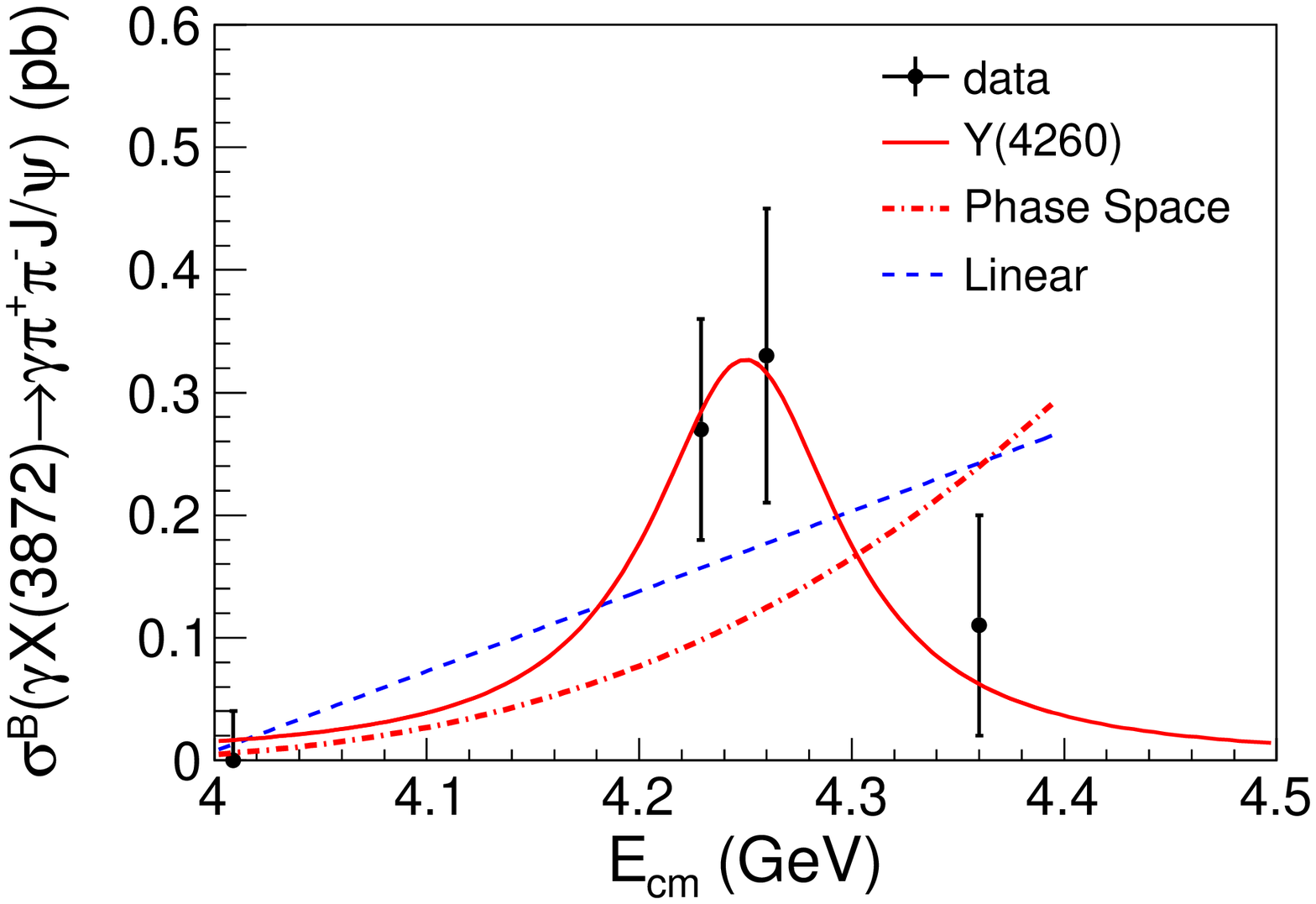}
\caption{Left panel: fit the $M(\ppjpsi)$ distribution observed at
BESIII. Dots with error bars are data, the curves are the best
fit. Right panel: fit to $\sigma^B[\EE\to \gamma\x]\times
\mathcal{B}[\x\to \ppjpsi]$ measured by BESIII with a $\y$
resonance (red solid curve), a linear continuum (blue dashed
curve), or an $E1$-transition phase space term (red dotted-dashed
curve). Dots with error bars are data.} \label{fit-mx}
\end{center}
\end{figure}

The $M(\ppjpsi)$ distribution (summed over all energy points), as
shown in Fig.~\ref{fit-mx}~(left), was fitted to extract the mass
and signal yield of $\x$. The ISR $\psp$ signal was used to
calibrate the absolute mass scale and to extract the resolution
difference between the data and a Monte Carlo (MC) simulation.
Figure~\ref{fit-mx} shows the fitted result: the measured mass of
$\x$ was $(3871.9\pm 0.7\pm 0.2)$~MeV/$c^2$. From a fit with a
floating width, we obtain a width of $(0.0^{+1.7}_{-0.0})$~MeV, or
less than 2.4~MeV at the 90\% C.L. The statistical significance of
$\x$ is $6.3\sigma$.

The Born-order cross-section was measured, and the results are
listed in Table~\ref{sec}. For 4.009 and 4.36~GeV data, because
the $\x$ signal is not significant, upper limits on the production
rates are given at the 90\% C.L. The measured cross-sections at
around 4.26~GeV are an order of magnitude higher than the NRQCD
calculation of continuum production~\cite{ktchao}, which may
suggest the $\x$ events come from resonance decays.

\begin{table}[htbp]
\caption{The product of the Born cross section $\sigma^{B}(\EE\to
\gamma \x)$ and $\mathcal{B}(\x\to \ppjpsi)$ at different energy
points. The upper limits are given at 90\% C.L.}
{\begin{tabular}{@{}cc@{}} \toprule
  $\sqrt{s}$~(GeV)  & $\sigma^B[\EE\to \gamma\x]
           \cdot\mathcal{B}(\x\to \ppjpsi)$~(pb) \\  \hline
  4.009 &  $0.00\pm 0.04\pm 0.01$ or $<0.11$  \\
  4.229 &  $0.27\pm 0.09\pm 0.02$  \\
  4.260 &  $0.33\pm 0.12\pm 0.02$  \\
  4.360 &  $0.11\pm 0.09\pm 0.01$ or $<0.36$ \\ \botrule
\end{tabular}\label{sec}}
\end{table}

The energy-dependent cross-sections were fitted with a $\y$
resonance (parameters fixed to PDG~\cite{pdg} values), linear
continuum, or $E1$-transition phase space ($\propto E^3_\gamma$)
term. Figure~\ref{fit-mx}~(right) shows all the fitted results,
which imply that $\chi^2/{\rm ndf}=0.49/3$ (C.L. = 92\%), 5.5/2
(C.L. = 6\%), and 8.7/3 (C.L. = 3\%) for a $\y$ resonance, linear
continuum, and phase space distribution, respectively. Thus, the
$\y$ resonance describes the data better than the other two
options.

These observations strongly support the existence of the radiative
transition process $\y\to \gamma\x$. The $\y\to \gamma\x$ process
could be another previously unseen decay mode of the $\y$
resonance. Together with the transitions to the charged
charmonium-like state $\zc$~\cite{zc3900,belley_new,seth_zc}, this
suggests that there might be some commonality in the nature of
$\x$, $\y$, and $\zc$, and so the model developed to interpret any
one of them should also consider the other two. As an example, the
authors of Ref.~\cite{model} integrated these states into a
molecular picture to calculate $\EE\to \gamma \x$ cross-sections.

Combining the above with the $\EE\to \ppjpsi$ cross-section
measurement at $\sqrt{s}=4.26$~GeV from BESIII~\cite{zc3900}, we
obtain $\sigma^B[\EE\to \gamma\x]\cdot \BR[\x\to
\ppjpsi]/\sigma^B(\EE\to \ppjpsi) = (5.2\pm 1.9)\times 10^{-3}$,
under the assumption that $\x$ and $\ppjpsi$ are only produced
from $\y$ decays. If we take $\BR[\x\to \ppjpsi] =
5\%$~\cite{bnote}, then $\mathcal{R} = \frac{\BR[\y\to
\gamma\x]}{\BR(\y\to \ppjpsi)}\sim 0.1$.

\section{\boldmath Observation of $\psi(1\,^3D_2)$}

BESIII observed $X(3823)$ in the $e^+e^-\to \pi^+\pi^-X(3823) \to
\pi^+\pi^-\gamma\chi_{c1}$ process with a statistical significance
of $6.2\sigma$ in data samples at c.m. energies of
$\sqrt{s}=$4.23, 4.26, 4.36, 4.42, and 4.60~GeV~\cite{BES3x3823}.

\begin{figure}
\begin{center}
\includegraphics[height=5.5cm]{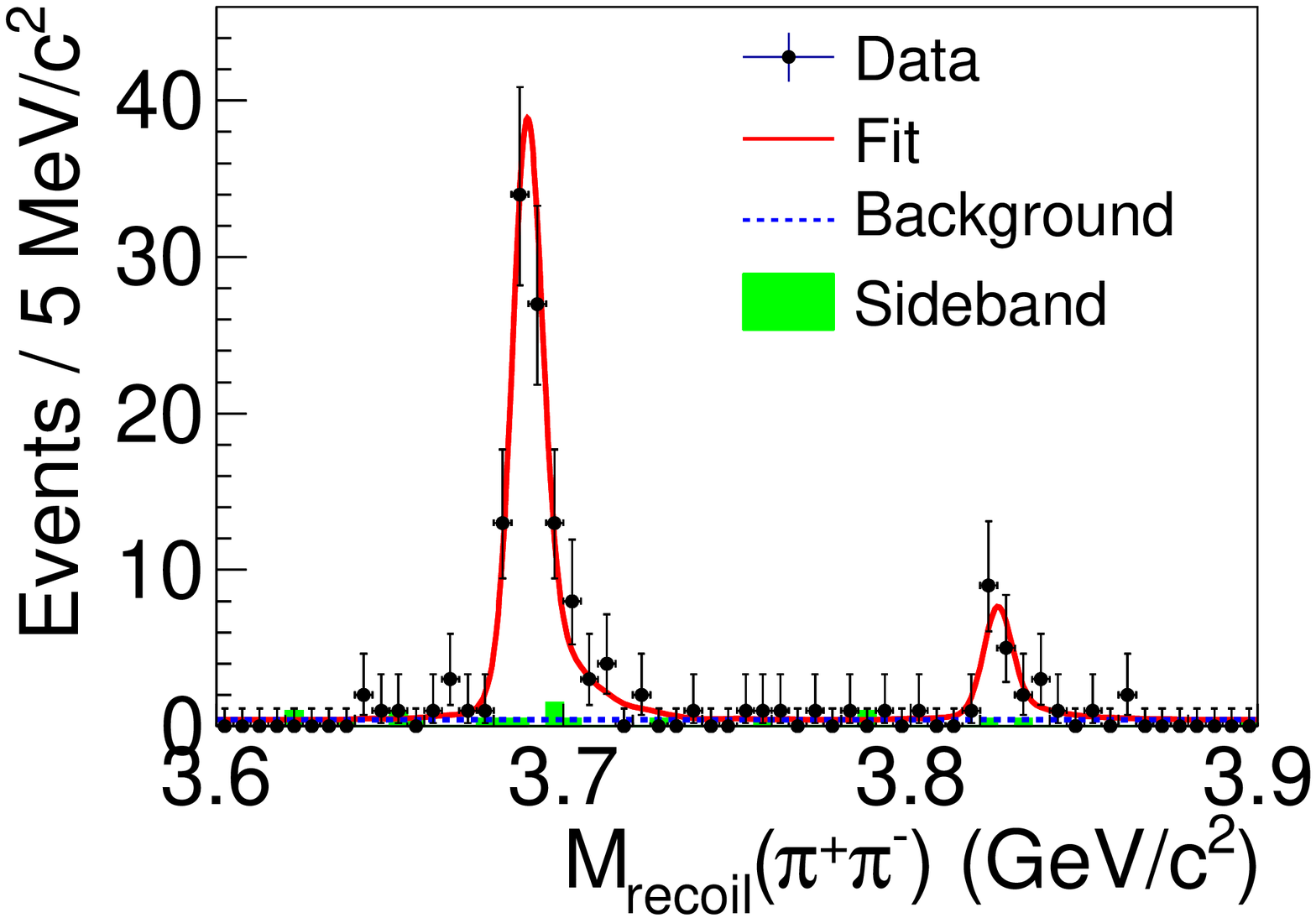}
\includegraphics[height=5.5cm]{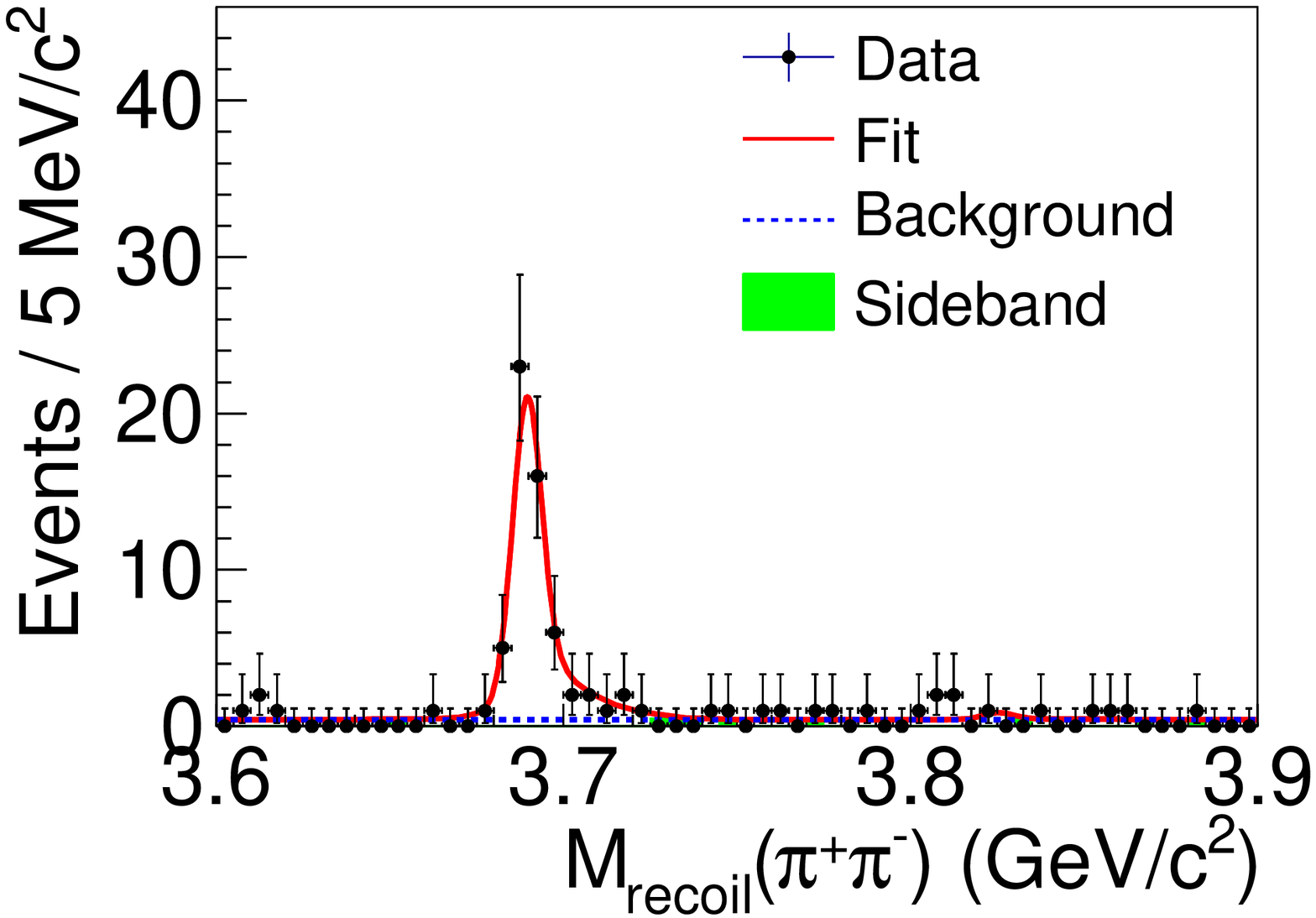}
\caption{Simultaneous fit to the $M_{\rm recoil}(\pp)$
distribution of $\gamma\chi_{c1}$ events (left) and
$\gamma\chi_{c2}$ events (right), respectively. Dots with error
bars are data, red solid curves are total fit, dashed blue curves
are background, and the green shaded histograms are $\jpsi$ mass
sideband events.} \label{X-fit}
\end{center}
\end{figure}

Figure~\ref{X-fit} shows the fitted results to $\pp$ recoil mass
distributions for events in the $\chi_{c1}$ and $\chi_{c2}$ signal
regions. The fit yields $19\pm 5$ $X(3823)$ signal events in the
$\gamma\chi_{c1}$ mode, with a  measured mass of $X(3823)$ of
$(3821.7\pm 1.3\pm 0.7)~{\rm MeV}/c^2$, where the first error is
statistical and the second systematic. For the $\gamma\chi_{c2}$
mode, no significant $X(3823)$ signal was observed, and an upper
limit on its production rate could be determined. The limited
statistics do not allow a measurement of the intrinsic width of
$X(3823)$. From a fit using the BW function (with a floating width
parameter) convolved with Gaussian resolution, it can be
determined that $\Gamma[X(3823)]<16$~MeV at the 90\% C.L.
(including systematic errors). This measurement agrees well with
the values found by Belle~\cite{belle-3d2}.

\begin{figure}
\begin{center}
\includegraphics[height=2.5in]{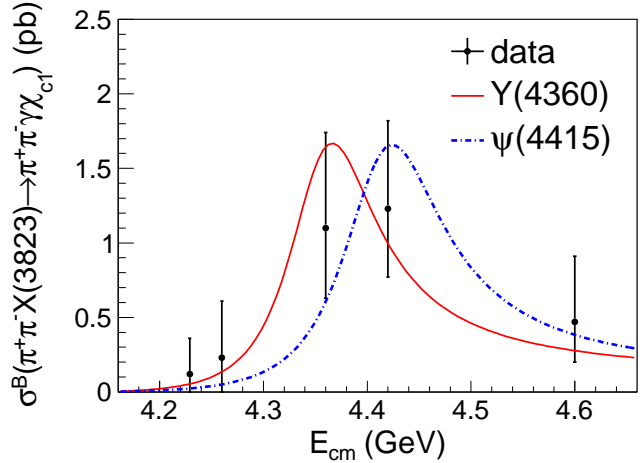}
\caption{Comparison of the energy-dependent cross sections of
$\sigma^B[\EE\to\pp X(3823)]\cdot
\mathcal{B}(X(3823)\to\gamma\chi_{c1})$ to the $Y(4360)$ and
$\psi(4415)$ line shapes. Dots with error bars (statistical only)
are data.  The red solid (blue dashed) curve shows a fit with the
$Y(4360)$ ($\psi(4415)$) line shape.} \label{fit-sec}
\end{center}
\end{figure}

The production cross-sections of $\sigma^{B}(\EE\to\pp
X(3823))\cdot \mathcal{B}(X(3823)\to \gamma\chi_{c1}$,
$\gamma\chi_{c2})$ were also measured at these c.m. energies. The
cross-sections of $\EE\to\pp X(3823)$ were fitted with the
$Y(4360)$ shape or the $\psi(4415)$ shape, with their resonance
parameters fixed to the PDG values~\cite{pdg}.
Figure~\ref{fit-sec} shows the fitted results, which give the
Kolmogorov--Smirnov statistic $D^{\rm H1}_{5,{\rm obs}}=0.151$ for
the $Y(4360)$ hypothesis (H1) and $D^{\rm H2}_{5,{\rm obs}}=0.169$
for the $\psi(4415)$ hypothesis (H2), based on the
Kolmogorov--Smirnov test~\cite{Kolmogorov}. Thus, both the
$Y(4360)$ and $\psi(4415)$ hypotheses ($D^{\rm H1}_{5,{\rm obs}},
D^{\rm H2}_{5,{\rm obs}}<D_{5,0.1}=0.509$) are accepted at a 90\%
C.L.

The $X(3823)$ resonance is a good candidate for the
$\psi(1\,^3D_2)$ charmonium state. According to potential
models~\cite{potential}, the $D$-wave charmonium states are
expected to be within a mass range of 3.82--3.85~GeV. Among
these, the $1\,^1D_2\to \gamma\chi_{c1}$ transition is forbidden
because of C-parity conservation, and the amplitude for
$1\,^3D_3\to \gamma\chi_{c1}$ is expected to be
small~\cite{barnes}. The mass of $\psi(1\,^3D_2)$ is in the
$3.810\sim 3.840$~GeV/$c^2$ range predicted by several
phenomenological calculations~\cite{3d2-mass}. In this case, the
mass of $\psi(1\,^3D_2)$ was above the $D\bar{D}$ threshold but
below the $D\bar{D}^*$ threshold. Because $\psi(1\,^3D_2)\to
D\bar{D}$ violates parity, $\psi(1\,^3D_2)$ is expected to be
narrow, in agreement with the observation, and $\psi(1\,^3D_2)\to
\gamma\chi_{c1}$ is expected to be a dominant decay
mode~\cite{3d2-mass, ratio}. From the cross-section measurement,
we obtain the ratio $\frac{\mathcal{B}[X(3823)\to
\gamma\chi_{c2}]}{\mathcal{B}[X(3823)\to \gamma\chi_{c1}]}<0.42$
(where systematic uncertainty cancels) at the 90\% C.L., which
also agrees with expectations for the $\psi(1\,^3D_2)$
state~\cite{ratio}.

\section{\boldmath Search for $\EE\to \gamma \chi_{cJ}$ and $\EE\to \gamma Y(4140)$}

Using data samples collected at $\sqrt{s}$ = 4.009, 4.23, 4.26,
and 4.36~GeV, BESIII searched for $e^+e^-\to \gamma\chi_{cJ}$ $(J
= 0, 1, 2)$ with the subsequent decay $\chi_{cJ}\to \gamma J/\psi$
and $J/\psi \to \mu^+\mu^-$. Evidence for the $e^+e^-\to
\gamma\chi_{c1}$ and $e^+e^-\to \gamma\chi_{c2}$ processes was
observed  with a statistical significance of 3.0$\sigma$ and
3.4$\sigma$, respectively~\cite{hezy}. No evidence for $\EE\to
\gamma\chi_{c0}$ was observed. The corresponding Born
cross-sections of $e^+e^- \to\gamma\chi_{cJ}$ at different c.m.
energies were calculated. Under the assumption that $\chi_{cJ}$
signals were absent, the upper limits on the Born cross-sections
were calculated at the 90\% C.L. These upper limits on the Born
cross-section of $\EE\to \gamma\chi_{cJ}$ are compatible with the
theoretical prediction from the NRQCD calculation~\cite{ktchao}.

BESIII searched for $Y(4140)$ via $\EE \to \gamma \phi\jpsi$ at
$\sqrt{s} = 4.23$, 4.26, and 4.36~GeV, but no significant
$Y(4140)$ signal was observed in any of the data
samples~\cite{BES3_Y4140}. The upper limits of the product of the
cross-section and branching fraction $\sigma[\EE \rightarrow
\gamma Y(4140)] \cdot \mathcal{B}(Y(4140) \rightarrow \phi\jpsi)$
at the 90\% C.L. were estimated to be 0.35, 0.28, and 0.33~pb at
$\sqrt{s} = 4.23$, 4.26, and 4.36~GeV, respectively.

These upper limits can be compared with the $X(3872)$ production
rates~\cite{BES3x}, which were measured from the same data
samples~(see Sec.~\ref{sec:x3872}). The latter were $\sigma[\EE
\to \gamma X(3872)] \cdot \mathcal{B}(X(3872) \rightarrow  \pi^{+}
\pi^{-} \jpsi) = [0.27 \pm 0.09({\rm stat}) \pm 0.02({\rm
syst})]$~pb, $[0.33 \pm 0.12({\rm stat}) \pm 0.02({\rm
syst})]$~pb, and $[0.11 \pm 0.09({\rm stat}) \pm 0.01({\rm
syst})]$~pb at $\sqrt{s} = 4.23, 4.26$, and $4.36$~GeV,
respectively, which are of the same order of magnitude as the
upper limits of $\sigma[\EE \rightarrow \gamma Y(4140)] \cdot
\mathcal{B}(Y(4140) \rightarrow \phi\jpsi)$ at the same energy.

The branching fraction $\mathcal{B}(Y(4140)\to \phi \jpsi)$ has
not previously been measured. Using the partial width of $Y(4140)
\to \phi\jpsi$ calculated under the molecule
hypothesis~\cite{y4140exp:4} and the total width of the $Y(4140)$
measured by CDF~\cite{y4140b}, the branching fraction was
estimated to be approximately 30\%. A rough estimation for
$\mathcal{B}(X(3872) \rightarrow \pi^{+} \pi^{-} \jpsi)$ is
5\%~\cite{bnote}. Combining these numbers, the ratio ${\sigma[\EE
\to\gamma Y(4140)]}/{\sigma[\EE \to\gamma X(3872)]}$ can be
estimated to be of the order of 0.1 or less at $\sqrt{s} = 4.23$
and 4.26~GeV.

\section{\boldmath Structures in $\EE\to {\rm charmonium}+{\rm hadrons}$}

\subsection{\boldmath Observation of $\EE\to \omega\chi_{c0}$}

Based on data samples collected between $\sqrt{s}=4.21$ and
4.42~GeV, the  $e^+e^-\to \omega\chi_{c0}$ process was observed at
$\sqrt{s}=4.23$ and 4.26~GeV for the first
time~\cite{BES3_omegachic0}, and the Born cross-sections were
measured to be $(55.4\pm 6.0\pm 5.9)$ and $(23.7\pm 5.3\pm
3.5)$~pb, respectively. For other energy points, no significant
signals were found, and upper limits on the cross-section at the
90\% C.L. were determined.

\begin{figure}[htbp]
\begin{center}
\includegraphics[width=0.6\textwidth]{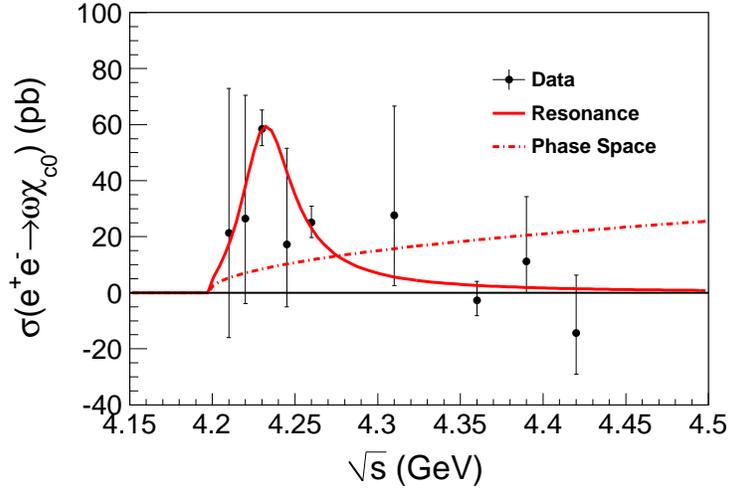}
\caption{Fit to $\sigma(e^+e^-\to\omega\chi_{c0})$ with a
resonance (solid curve), or a phase space term (dot-dashed curve).
Dots with error bars are the dressed cross sections. The
uncertainties are statistical only. } \label{CS-BW-float}
\end{center}
\end{figure}

The data reveal a sizable $\omega\chi_{c0}$ production at around
4.23~GeV/$c^2$, as predicted in Ref.~\cite{omg-chi0}. By assuming
the $\omega\chi_{c0}$ signals come from a single resonance, the
$\Gamma_{ee} \mathcal{B} (\omega\chi_{c0})$, mass, and width of
the resonance were determined to be $(2.7\pm 0.5\pm 0.4)$~eV,
$(4230\pm 8\pm 6)$~MeV/$c^2$, and $(38\pm 12\pm 2)$~MeV,
respectively (the fit is shown in Fig.~\ref{CS-BW-float}). The
parameters are consistent with those of the narrow structure in
the $\EE\to \pp\hc$ process~\cite{y4220_ycz}, but are not
consistent with those obtained by fitting a single resonance to
the $\pi^+\pi^-J/\psi$ cross-section~\cite{babary} (with a mass of
$(4259\pm 8^{+2}_{-6})~{\rm MeV}/c^2$ and a width of $(88\pm
23^{+6}_{-4})~{\rm MeV}$). This suggests that the observed
$\omega\chi_{c0}$ signals are unlikely to originate from
$Y(4260)$.

The $e^+e^-\to \omega\chi_{c1,2}$ channels were also sought, but
no significant signals were observed; upper limits at the 90\%
C.L. on the production cross-sections were determined. The very
small measured ratios of $e^+e^-\to\omega\chi_{c1,2}$
cross-sections to those for $e^+e^-\to\omega\chi_{c0}$ are
inconsistent with the prediction in Ref.~\cite{Ratio}.

\subsection{\boldmath Measurement of $\EE\to \eta\jpsi$}

Using data samples collected at energies of 3.81--4.60~GeV, BESIII
analyzed $e^{+}e^{-} \to \eta J/\psi$~\cite{etajpsi}.
Statistically significant $\eta$ signals were observed, and the
corresponding Born cross-sections were measured. In addition, a
search for the $e^{+}e^{-} \to \pi^0 J/\psi$ process observed no
significant signals, and upper limits at the 90\% C.L. on the Born
cross-section were set.

%%%%%%%%%%%%%%%%%%%%
\begin{figure}[htbp]
\begin{center}
\begin{overpic}[width=7.0cm,height=5.0cm,angle=0]{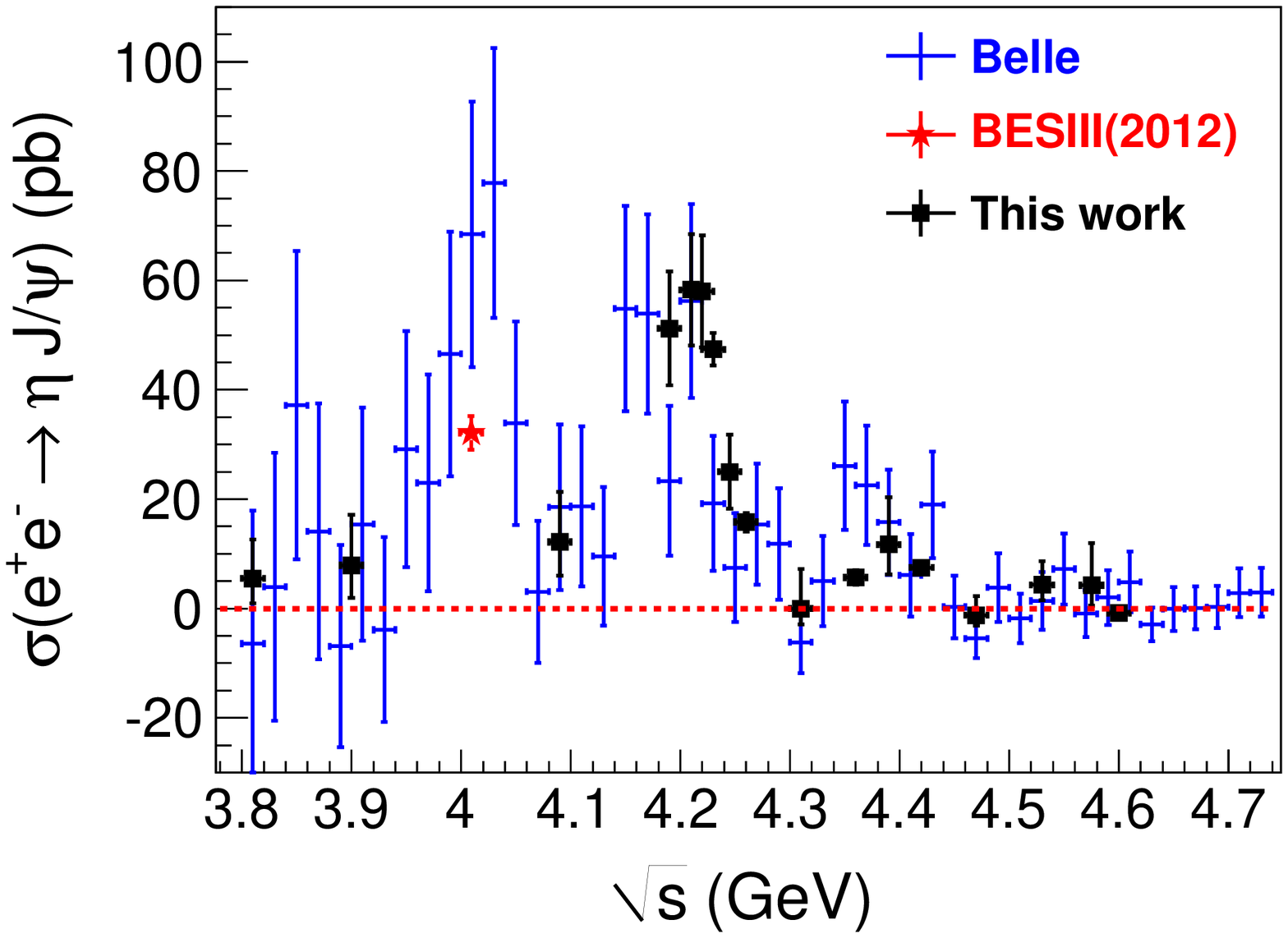}
\put(19,57){\large\bf (a)}
\end{overpic}
\begin{overpic}[width=7.0cm,height=5.0cm,angle=0]{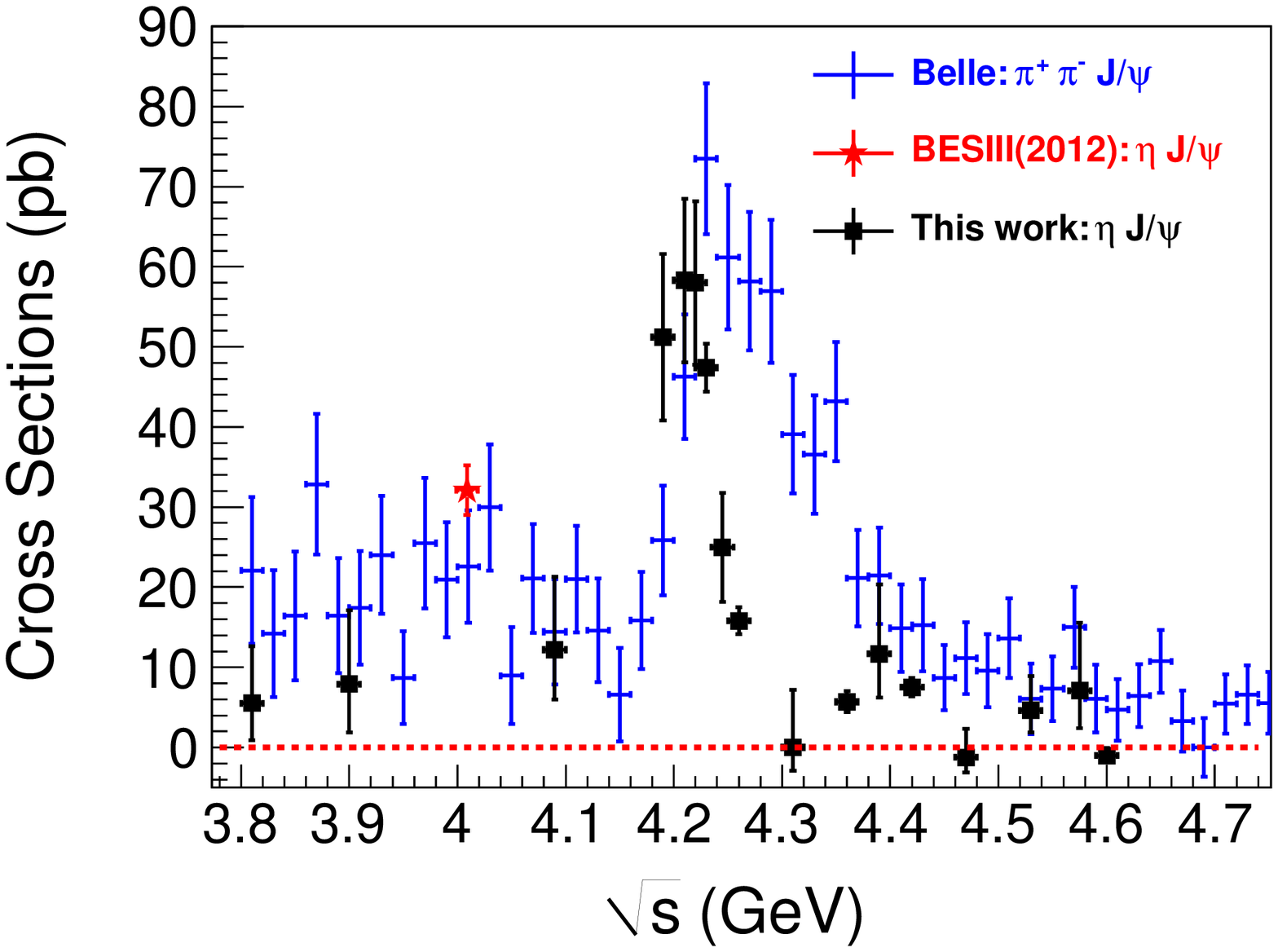}
\put(19,57){\large\bf (b)}
\end{overpic}
\end{center}
\caption{A comparison of the measured Born cross sections of
$e^{+}e^{-} \to \eta J/\psi$ to those of the previous
measurements~\cite{4040,belle} (a), and to those of $e^{+}e^{-}\to
\pi^{+}\pi^{-} J/\psi$ from Belle~\cite{belley_new}. In these two
plots, the black square dots and the red star dots are the results
of $\eta J/\psi$ obtained from BESIII. The blue dots are results
of $\eta J/\psi$ (a) and $\pi^{+}\pi^{-} J/\psi$ (b) from Belle.
The errors are statistical only for Belle's results, and are final
combined uncertainties for BESIII's results.}\label{BES_BELLE}
\end{figure}
%%%%%%%%%%%%

A comparison of the Born cross-sections $\sigma(e^{+}e^{-} \to
\eta J/\psi)$ in this measurement to previous
results~\cite{4040,belle} is shown in Fig.~\ref{BES_BELLE} (a),
indicating  very good agreement. The measured Born cross-sections
were also compared to those of $e^{+}e^{-} \to \pi^{+}\pi^{-}
J/\psi$ obtained from the Belle experiment~\cite{belley_new}, as
shown in Fig.~\ref{BES_BELLE} (b). Different line shapes can be
observed in these two processes, indicating that the production
mechanism of  $\eta J/\psi$ differs from that of
$\pi^{+}\pi^{-}J/\psi$ in the vicinity of $\sqrt{s}$ =
4.1--4.6~GeV. This could indicate the existence of a rich spectrum
of the $Y$ states in this energy region with different coupling
strengths to various decay modes.

\section{\boldmath More data for $\xyz$ study}

As shown in previous sections, BESIII has achieved a lot in the
study of the $\xyz$ states and the conventional charmonium states.
However, there are more questions to be answered with the
currently available data and, more importantly, with the data
samples that BESIII is able to collect in the next few years.

A few topics that  need to be studied with more data are listed
below.
\begin{itemize}
\item In the $X$ sector:
  \begin{itemize}
  \item Where are the $\x$ and $\psi(1\,^3D_2)$ coming from,
  resonance decays or  continuum production?
  \item Can the other $X$ states, such as $\xyz(3940)$, $X(3915)$,
  $X(4140)$ ($Y(4140)$), and $X(4350)$ be produced in a similar way?
  \item Can the charmonium 2P, 3P states and the S-wave spin-singlet
  states (3S, 4S, 5S) be observed in radiative transitions?
  \end{itemize}
\item In the $Y/\psi$ sector:
  \begin{itemize}
  \item Is the $\y$ structure a single resonance, or does it have a
  more complicated sub-structure? Is $Y(4008)$ a real resonance?
  \item What are the other decay modes of  $Y(4360)$?
  \item What is hidden in the $\EE\to \pphc$ line shape?
  \item Is the $Y(4660)$ observed in $\EE\to \pp\psp$
  the same as the $Y(4630)$ observed in
  $\EE\to \Lambda_c^+ \Lambda_c^-$~\cite{belle_y4630}?
  \item What is the correlation between charm production ($\EE\to$
  open charm final states) and charmonium production?
  \item Where is the vector charmonium $3D$ state?
  \item Are there charmonium states between $\psi(4160)$ and
  $\psi(4415)$, and/or between $\psi(4415)$ and $Y(4660)$?
  \item Can the vector charmonium hybrid state be observed~\cite{ccg_lqcd}?
  \end{itemize}
\item In the $Z$ sector:
  \begin{itemize}
  \item Are the $Z_c$ states produced from resonance decays or
  from continuum production?
  \item Is there a $Z_{cs}$ state decaying into $K^\pm\jpsi$ or
  $D_s^-D^{*0}+c.c.$, $D_s^{*-}D^0+c.c.$?
  \item Are there more $Z_c$ and $Z_{cs}$ states?
  \end{itemize}
\item In the $C$ sector:
  \begin{itemize}
  \item Can $D_{s0}(2317)$ be produced and studied?
  \item Can the other excited charmed mesons be produced and
  studied with high-energy data?
  \end{itemize}
\end{itemize}

BESIII is going to collect about 3~fb$^{-1}$ of data in the
vicinity of $\psi(4160)$ to study the $D_s$ decay properties in
2015--2016 run. These data can be used to answer some of the
questions listed above. A few specific topics are listed here.
\begin{enumerate}
\item In $\EE\to \ppjpsi$, there is a dip at around 4.17~GeV and a
sharp increase at around 4.23~GeV~\cite{belley_new}. The data may
help to identify where the turning point is located.

\item For the $\EE\to \omega\chi_{c0}$
process~\cite{BES3_omegachic0}, there are few data points close to
the threshold to support the claim of a narrow structure.

\item For the $\EE\to \kk\jpsi$ process, it seems there is a
structure at around 4.2~GeV~\cite{belle_kkjpsi}.

\item In the $\EE\to \eta\jpsi$ process, the line shape differs
from that of $\EE\to \ppjpsi$, with a peak at around 4.2~GeV, but
there are no data points on the left side.

\item In $\EE\to \eta\hc$, is there a structure close to the
threshold? What does the $\EE\to \etap\jpsi$ line shape look like
close to the threshold?

\item Are the $\EE\to \gamma \chi_{cJ}$ signals from $\psi(4160)$
or $Y(4260)$, or something else~\cite{hezy}?

\item The search for $F$-wave charmonium states via $\psi(4160)$
decays.

\item The search for $\gamma \chi_c(2P)$, $\gamma \xyz(3940)$,
$\gamma X(3915)$ from $\psi(4160)$ decays.

\end{enumerate}

More data are always better, but as we need to collect data at
many points and the collection period is limited, we need to
define lower limits for the number of data points and  the
luminosity needed at each point to ensure meaningful measurements
of the physical quantities of interest.

As we require high-precision cross-sections for all of the open
charm modes and the cross-sections of some hadronic transition
modes to fit the resonant parameters of the vector states, we
require sufficient data points to give the excitation curve of any
of the existing vector resonances and possible hidden structures
in the full energy range.

Some special issues need to be considered: (1) the thresholds of
all the open charm and charmonium + hadron final states; (2)
possible energy regions where large interference effects are
expected; (3) energy regions where narrow structures are expected;
(4) energy regions that were not well explored before; and (5) the
beam energy spread.

From previous calculations and measurements, it is known that the
energy spread of BEPCII at energy ranges of 3.8--5.0~GeV is around
1.4--2.0~MeV. In this case, we would not use energy steps finer
than three times the energy spread. That is, unless very
necessary, we would not take data at energy points less than 5~MeV
from an existing data point. Thus, we would take data with 5~MeV
steps in the energy regions where narrow structures or dramatic
effects are expected, such as the thresholds for open charm or
hidden charm final states, and the low mass shoulder of $Y(4260)$,
where interference effects have been reported~\cite{belley_new}.
Otherwise, 10~MeV steps will be used.

The limit on the data size is set according to the precision of
the hadronic transition modes, which typically have cross-sections
of a few to a few tens of picobarns. From the analyses in previous
sections, we find that an integrated luminosity of 500~pb$^{-1}$
is needed to reach a reasonably high precision for most modes of
interest. To search for the $\xyz$ and charmonium states via
radiative transition, a data sample of at least 500~pb$^{-1}$ is
also needed to reach a $5\sigma$ level observation of a signal if
the production rate is $1\times 10^{-4}$ or higher. A detailed MC
study of the precision that can be reached or the exact luminosity
needed for each energy point is necessary, as the background level
may be very different at different resonant peaks.

With the above principles in mind, we propose to collect data
samples at about 60 c.m. energies from 4.0~GeV to the maximum
energy that BEPCII can reach (currently 4.6~GeV) in 10~MeV steps.
Around 500~pb$^{-1}$ at each energy will be necessary for a
comprehensive study of the $\xyz$ and charmonium states. As BESIII
has already accumulated about 5~fb$^{-1}$~(see
Table~\ref{ecm_lum_xyz}), another 25~fb$^{-1}$ data should be
accumulated. This will take about 5 years at
BEPCII~\cite{BESIII_YB}.

\section{\boldmath $\xyz$ study in future experiments}

\begin{figure}[htbp]
\begin{center}
\includegraphics[width=0.6\textwidth]{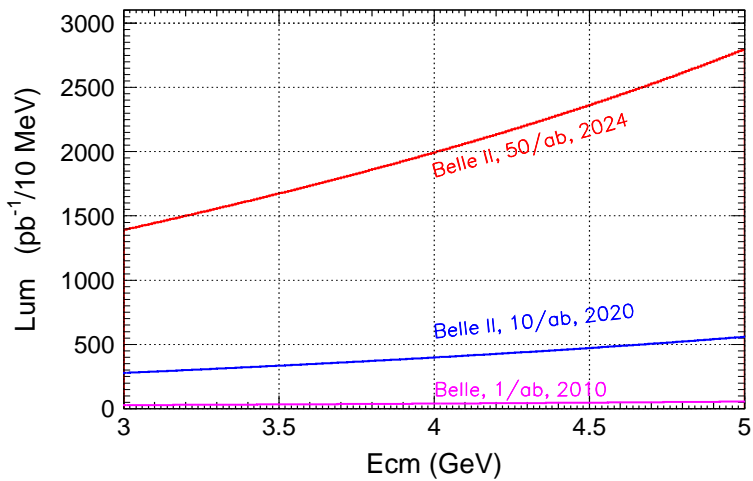}
\caption{Effective luminosity at low energy in the Belle and
Belle-II $\Upsilon(4S)$ data samples.} \label{lum_belle2}
\end{center}
\end{figure}

Belle-II~\cite{belle2} will start collecting data in 2018, and
will accumulate 50~ab$^{-1}$ data at the $\Upsilon(4S)$ peak by
2024. These data samples can be used to study the $\xyz$ and
charmonium states in many different ways~\cite{PBFB}, among which
ISR can produce events in the same energy range covered by BESIII.
Figure~\ref{lum_belle2} shows the effective luminosity at BEPCII
energy in the Belle-II data samples. We can see that, for
10~ab$^{-1}$ Belle-II data, we have about 400--500~pb$^{-1}$ data
for every 10~MeV in the range 4--5~GeV, comparable to the data
sample proposed at BESIII in the previous section. Of course, the
ISR analyses have a lower efficiency than in direct $\EE$
collisions because of the extra ISR photons and the boost given to
events along the beam direction. Even taking these effects  into
account, the full Belle-II data sample, which corresponds to about
2,000--2,300~pb$^{-1}$ data for every 10~MeV from  4--5~GeV, will
result in similar statistics for modes like $\EE\to \ppjpsi$.
Belle-II has the advantage that data at different energies will be
accumulated at the same time, making the analysis much simpler
than at BESIII at 60 data points. In addition, Belle-II can
produce events above 4.6~GeV, which is currently the maximum
energy of BEPCII. Possible upgrades to increase the maximum c.m.
energy of BEPCII will obviously expand the physical possibilities
of BESIII.

The HIEPA project~\cite{HIEPA} being discussed at this workshop
will improve the aforementioned studies in many aspects,
particularly with c.m. energies of up to 7~GeV and luminosity
improvements of a factor of 100. These will allow a finer scan in
the full energy region with more integrated luminosity. This will
enable a better understanding of all the studies listed in this
article.

\section{Conclusion}

With the world's largest data samples at energies of 3.8--4.6~GeV,
the BESIII experiment made a significant contribution to the study
of the charmonium and $\xyz$ states. To further strengthen such
studies, BESIII may collect more data from 4.0--4.6~GeV (or even
higher, if possible). These data will be complementary to the
Belle-II study, with many other production mechanisms. The HIEPA
project may enable a systematic understanding of the nature of the
$\xyz$ and charmonium states.

\acknowledgments

This work is supported in part by National Natural Science
Foundation of China (NSFC) under contract Nos. 11235011 and
11475187; the Ministry of Science and Technology of China under
Contract No. 2015CB856701, and the CAS Center for Excellence in
Particle Physics (CCEPP).

\end{document}